\documentclass[conference]{IEEEtran}
\IEEEoverridecommandlockouts
\usepackage{cite}
\usepackage{amsmath,amssymb,amsfonts}
\usepackage{algorithmic}
\usepackage{graphicx}
\usepackage{caption}
\usepackage{fancyhdr}
\usepackage{subcaption}
\usepackage{textcomp}
\usepackage{authblk}
\usepackage{xcolor}
\usepackage{url}
\def\BibTeX{{\rm B\kern-.05em{\sc i\kern-.025em b}\kern-.08em
    T\kern-.1667em\lower.7ex\hbox{E}\kern-.125emX}}
\begin{document}

\title{ML-Based Feedback-Free Adaptive MCS Selection for Massive Multi-User MIMO}
\author{Qing An, Mehdi Zafari, Chris Dick, Santiago Segarra,  Ashutosh Sabharwal, Rahman Doost-Mohammady
\thanks{Qing An, Mehdi Zafari, Santiago Segarra, Ashutosh Sabharwal and Rahman Doost-Mohammady are with the Department of Electrical and Computing Engineering, Rice University, Houston, TX, USA. E-mail:\{qa4, mz52, segarra, ashu, doost\} @rice.edu, Chris Dick is with NVIDIA Corporation, Santa Clara, CA, USA. E-mail: cdick@nvidia.com }}

\maketitle
\thispagestyle{fancy}        
\fancyhead{}                     
\lhead{This work has been accepted to 2023 Asilomar Conference on Signals, Systems, and Computers. \\
Copyright may be transferred without notice, after which this version may no longer be accessible.}     
\chead{}
\rhead{}
\lfoot{}
\cfoot{\quad}  
\rfoot{}
\renewcommand{\headrulewidth}{0pt}  
\renewcommand{\footrulewidth}{0pt}
\pagestyle{empty}


\begin{abstract}
As wireless communication systems strive to improve spectral efficiency, there has been a growing interest in employing machine learning (ML)-based approaches for adaptive modulation and coding scheme (MCS) selection. In this paper, we introduce a new adaptive MCS selection framework for massive MIMO systems that operates without any feedback from users by solely relying on instantaneous uplink channel estimates. Our proposed method can effectively operate in multi-user scenarios where user feedback imposes excessive delay and bandwidth overhead. To learn the mapping between the user channel matrices and the optimal MCS level of each user, we develop a Convolutional Neural Network (CNN)-Long Short-Term Memory Network (LSTM)-based model and compare the performance with the state-of-the-art methods. Finally, we validate the effectiveness of our algorithm by evaluating it experimentally using real-world datasets collected from the RENEW massive MIMO platform.
\end{abstract}



\section{Introduction}
\label{sec:intr}
The growing demand for high-speed wireless connectivity has led to the adoption of advanced wireless communication technologies such as massive multi-user multiple-input multiple-output (MIMO) systems. Massive MIMO enhances the spectral efficiency (SE) in wireless networks by leveraging a large number of service antennas to serve a large number of users on the same time-frequency channel resource. However, rate adaptation by selecting the appropriate transmission link parameters, including modulation and coding scheme (MCS), to maximize spectral efficiency remains a challenging task, particularly in dynamic wireless environments where channel conditions can vary rapidly. Adaptive modulation and coding (AMC) has been studied as a critical block in MAC/PHY layers of wireless standards, and various algorithms have been developed to enable the dynamic adjustment of MCS levels in response to changing channel conditions~\cite{amc_ref}.

 In a time division duplex (TDD)-based massive MIMO system, base stations (BSs) can acquire channel state information (CSI) through uplink training thanks to reciprocity. The 5G standard defines several pilot patterns for channel estimation, where the appropriate pilot pattern and pilot period will be determined by the network operator depending on the configuration. It also follows in the standard that user equipment (UE) has to report a channel quality indicator (CQI) to the serving BS to be used to decide the MCS in the next transmission slot. In current standard implementations, the period of the channel estimation is typically significantly shorter than the CQI feedback period~\cite{survey}.~

Recently, machine learning (ML) has emerged as a promising methodology applied across diverse aspects of wireless communication~\cite{10247079,online_dl,dqn_amcs_sdm}, encompassing adaptive modulation and coding. Various techniques have been proposed to utilize the CQI feedback and CSI information as inputs for different ML models, allowing for efficient learning of the optimal mapping between channel conditions and MCS levels. An online deep learning algorithm for adaptive modulation and coding in massive MIMO systems is proposed in \cite{online_dl}, which leverages self-learning in a deep neural network (DNN) model. In \cite{dqn_amcs_sdm}, authors proposed a deep Q-network (DQN)-based joint adaptive scheduling algorithm of MCS and space division multiplexing (SDM) for 5G massive MIMO. A hybrid data-driven and model-based ML approach for link adaptation is proposed in \cite{link_adaptation_hybrid}, which leverages CSI history to optimally select the MCS. These methods rely on conforming to the standard requirements and presume that every user will provide CQI feedback. Furthermore, most of the proposed methods in the literature provide performance analysis through only limited numerical results derived from simulated channels.

In massive multi-user MIMO systems, where a large number of users are scheduled on the same time-frequency resource, and an even larger number of users are connected to each serving BS, having feedback from each user causes the delay to grow significantly, making the user feedback prohibitive due to imposing an excessive delay. Nevertheless, CQI acquisitions for beamformed users are independent such that each individual user may not know how many other MU-MIMO layers are being adopted. Therefore, CQI feedback lacks context, which is even more severe in massive MIMO networks. Moreover, the CQI feedback period is typically long so the downlink CQI of the previous slot can become outdated to be used as a reference to do MCS selection for the next slot, especially in significantly varying channel environments (e.g. high mobility). This motivates the need for an adaptive MCS selection framework that relies only on the instantaneous uplink channel estimates, removing the need for any feedback from users.

Additionally, sequential correlation is a pivotal factor in AMC strategies. To enhance performance, Outer Loop Link Adaptation (OLLA)~\cite{olla} has been introduced as an improvement over traditional Inner Loop Link Adaptation (ILLA). Apart from Channel Quality Indicator (CQI) feedback, OLLA incorporates the acknowledgment (ACK)/negative acknowledgment (NACK) from previous time slots in the uplink control information. This inclusion of sequential data from prior slots can provide insightful channel condition context to AMC, especially in mobility scenarios. Recent works based on reinforcement learning (RL)~\cite{10024770} have explored the integration of CQI or CSI from multiple time slots into the state space to guide the model in capturing channel sequential correlations. Nonetheless, this approach results in an expanded state space, thereby posing challenges in convergence, especially as the network scales. Consequently, it is imperative to consider alternative models that are based on sequential dependencies to address this issue.

In this paper, we present an innovative ML-based framework for adaptive modulation and coding in massive MIMO networks. 
We introduce a feedback-free AMC technique that capitalizes on the CSI acquired through uplink sounding reference signals (SRS) to facilitate the selection of the MCS for the subsequent downlink transmission. 
We present a model based on Convolutional Neural Networks and Long Short-Term Memory networks (CNN-LSTM) designed to effectively learn the optimal mapping between instantaneous uplink CSI and the most suitable MCS for individual users. CNNs are adept at extracting spatial information from the channel matrix, particularly relevant in multi-user transmission scenarios with inter-user correlation. Additionally, LSTMs excel at processing sequential correlations across different transmission time slots. 
We demonstrate the effectiveness of the proposed algorithm through extensive experimentation with real-world datasets collected from the RENEW platform \cite{renew}, a programmable multi-cell massive MIMO base station deployed on the Rice University campus.
Our approach enables the adaptive adjustment of MCS levels based solely on channel estimates with high accuracy and thus significantly reduces feedback delays by obviating the requirement for user-provided CQI feedback. 
Apart from multi-user systems, the proposed algorithm is applicable to the single-user regime as well.






\section{Proposed Adaptive MCS Algorithm}
\label{sec:alg}
\subsection{Overview of Multi-User Feedback-Free AMC}
\begin{figure}[t]
      \centering
      \includegraphics[width=0.48\textwidth]{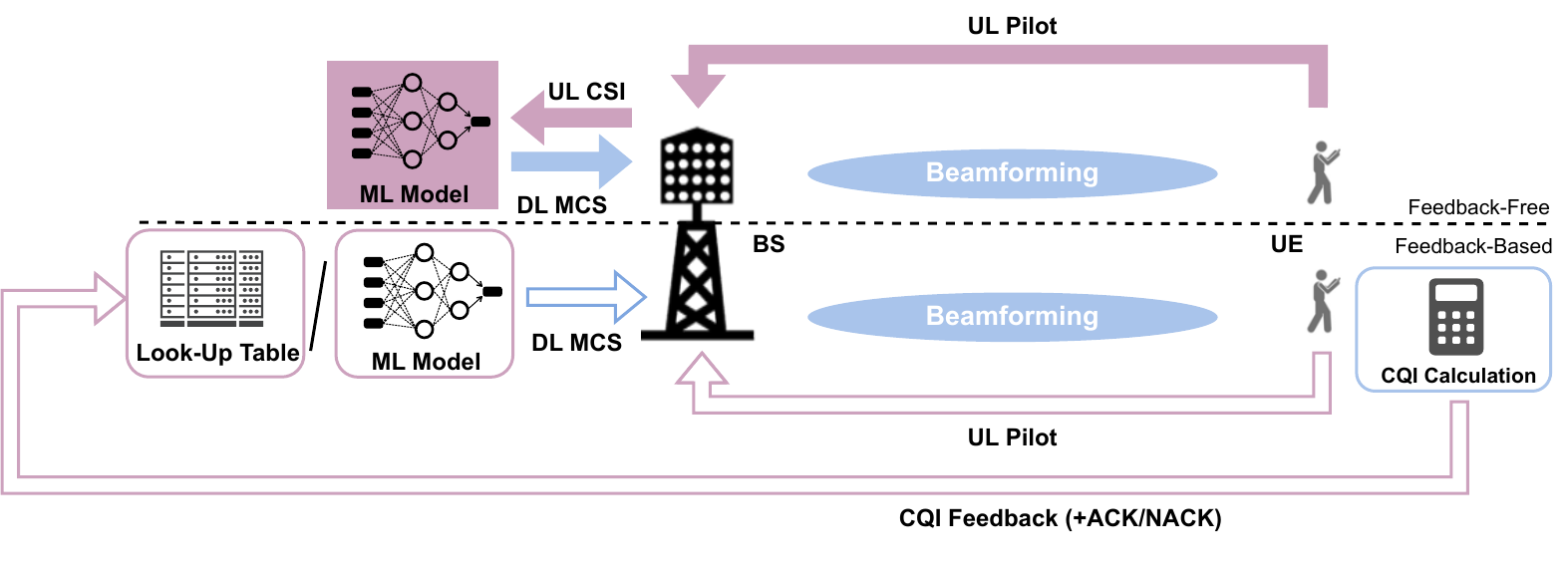}
      \caption{Comparison of a Feedback-Free and a Feedback-based Adaptive MCS selection scheme.} 
      \label{fig:ff}
\end{figure}

An illustration of a feedback-free adaptive MCS selection scheme compared to a feedback-based technique is provided in Fig.~\ref{fig:ff}. In a traditional feedback-based framework, UEs transmit CQI feedback to the base station via an uplink control channel, and the base station subsequently maps the received CQI values to specific MCS levels by referencing a predefined lookup table~\cite{3gpp.38.214}. Some existing literature~\cite{online_dl,dqn_amcs_sdm,10024770} also incorporates ML-based methods for this mapping process. In an effort to enhance the accuracy of this mapping, some approaches include Hybrid Automatic Repeat Request (HARQ) information (i.e. ACK/NACK) from previous time slots in the feedback signals, as the sequential dependency of channel condition plays a critical role in adaptive MCS selection, particularly in mobile scenarios.

Nevertheless, 
according to~\cite{3gpp.38.214}, CQI feedback typically consumes 4 bits per transmission, incurs up to an 8-ms delay, and happens every 8 time slot. In cases where radio resources are limited, and stringent time constraints exist, the feedback-based approach may struggle to adjust MCS levels promptly, leading to a degradation in system performance. In contrast, the feedback-free-based (FF-based) method leverages the uplink channel matrix, estimated through channel measurements using uplink pilots, to eliminate the overhead associated with bandwidth consumption. Channel measurements exhibit a significantly shorter periodicity compared to CQI feedback, ensuring the real-time adaptability of MCS.

\subsection{ML-based AMC Model}

Our FF-based AMC aims to predict the MCS for the next downlink transmission by employing an ML-based approach. We focus on wideband AMC, where the same MCS is assigned to the entire bandwidth. The input for our ML model is a 3D channel matrix, with dimensions corresponding to base station antennas, user equipment, and the real and imaginary parts of complex entries in the channel matrix. To perform AMC using the channel matrix, it is critical to use a model that can extract intricate inter-user correlation from the channel matrix, which is a key determining factor in the downlink channel quality of the users. Akin to the use of convolutional neural networks (CNNs) to extract hierarchical features in images, we also adopt a DenseNet-based CNN architecture~\cite{dense}, a variant of deep CNNs that solve the well-known vanishing gradients in CNNs using skip-connections.  

\begin{figure}[ht]
      \centering
      \includegraphics[width=0.48\textwidth]{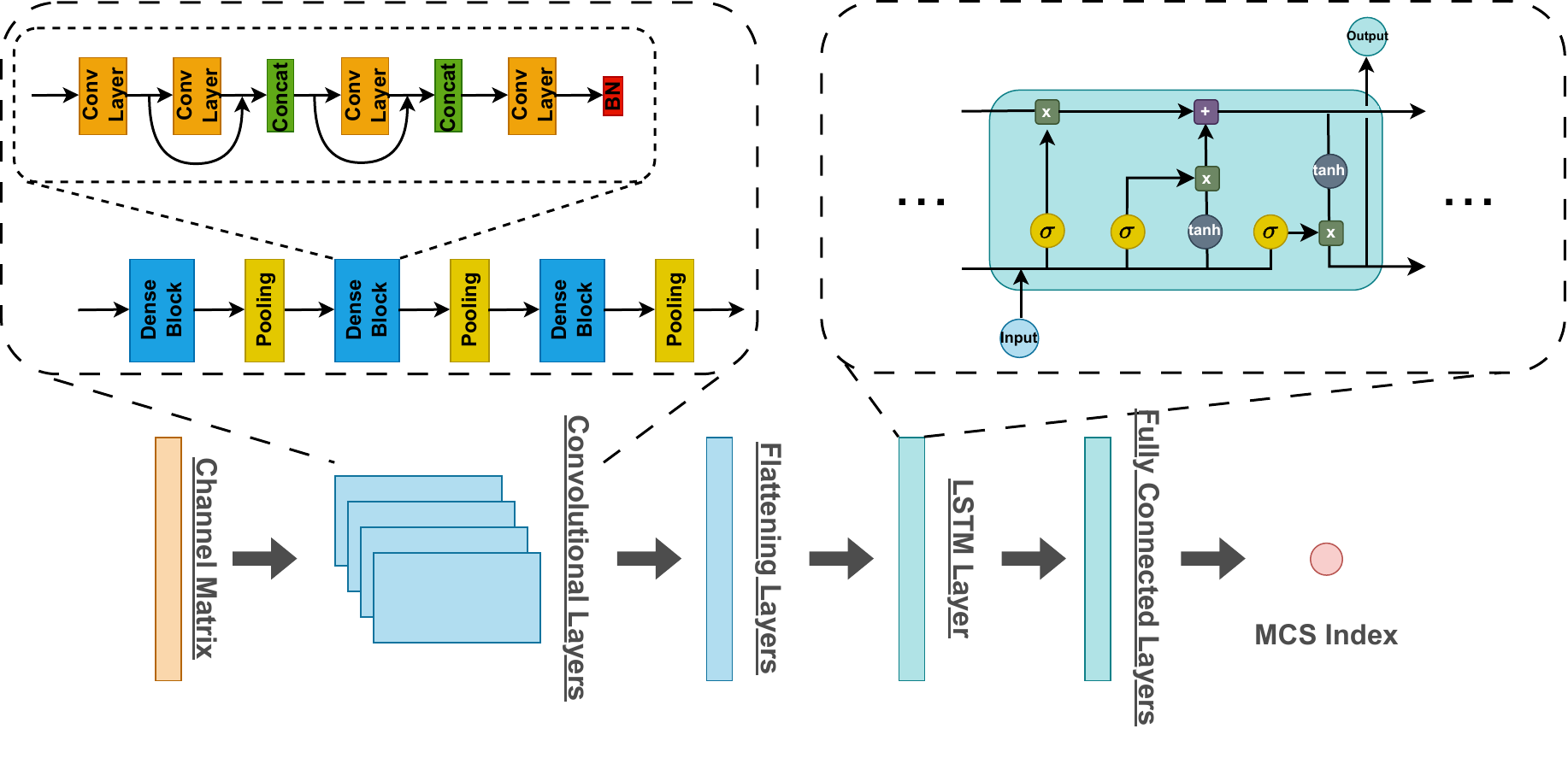}
      \caption{ML-based framework.} 
      \label{fig:ml}
\end{figure}

Another important factor in optimally predicting the MCS is the need to capture the temporal dependencies that exist between successive transmission slots as mentioned in Sec.~\ref{sec:intr}. 
Long Short-Term Memory (LSTM) networks~\cite{lstm}, a type of recurrent neural network (RNN)~\cite{rnn}, are known to excel in modeling sequential data and time-series forecasting. To capture the temporal correlation in channel data, we add an LSTM block to our overall architecture. 

Fig. \ref{fig:ml} illustrates our employed architectural framework, which includes three key components: a CNN-based segment, an LSTM-based segment, and three Fully Connected Layers (FCLs). The CNN employs \emph{Dense Blocks} each consisting of four convolutional layers, maintaining consistent feature map sizes and incorporating skip-connections via concatenation layers. Batch Normalization (BN) layers are added at the end of each Dense Block to mitigate overfitting. Average pooling layers are placed between Dense Blocks to reduce feature map dimensions. The \emph{flattening layer} transforms CNN outputs into a flat format for input to LSTM networks. The LSTM structure comprises three memory blocks with forget, input, and output gates, employing sigmoid functions to control information flow and tanh functions to adjust values. Ultimately, the FCLs combine features extracted by the CNN and LSTM to generate the MCS prediction.

\section{Performance Evaluation}
\label{sec:exp}

In this section, we assess our adaptive MCS framework by conducting a thorough evaluation. We gather simulation and over-the-air (OTA) data using Agora\cite{agora}, a real-time software implementation of massive MIMO PHY/MAC for both massive MIMO BS and users. Agora can be run in simulator mode where datasets from QUAsi Deterministic RadIo channel GenerAtor (QuaDRiGa)\cite{qua} can be replayed using a software channel simulator between Agora BS and Agora user software\cite{doost2022getmobile}. Agora can also be run on the RENEW massive MIMO BS and single-antenna clients that emulate UEs. We benchmark our approach against various state-of-the-art (SOTA) techniques, including heuristic and ML-based methods, to evaluate testing accuracy. This analysis demonstrates our model's effectiveness in selecting the optimal MCS.

\subsection{Experiment Setup}

\begin{figure}[ht!]
    \centering
    \begin{subfigure}[b]{0.15\textwidth}
    \centering
      \includegraphics[width=\textwidth]{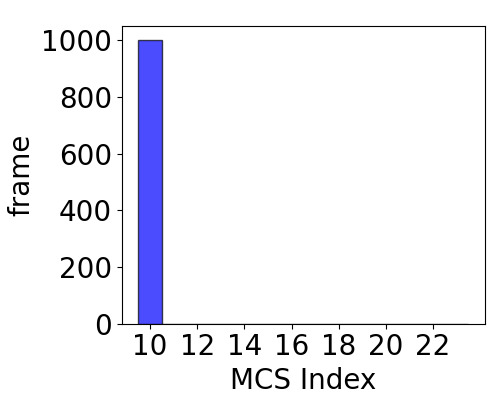}
      \caption{}
      \label{fig:c1}
    \end{subfigure}
    \begin{subfigure}[b]{0.15\textwidth}
    \centering
      \includegraphics[width=\textwidth]{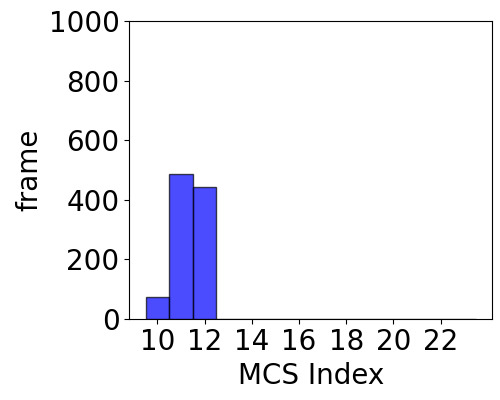}
      \caption{}
      \label{fig:c2}
    \end{subfigure}
    \begin{subfigure}[b]{0.15\textwidth}
    \centering
        \includegraphics[width=\textwidth]{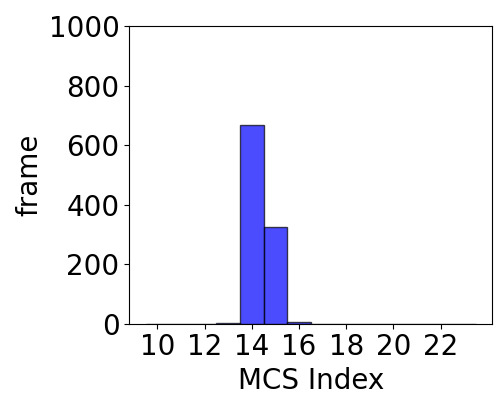}
        \caption{}
        \label{fig:uc}
    \end{subfigure}
    \begin{subfigure}[b]{0.15\textwidth}
    \centering
        \includegraphics[width=\textwidth]{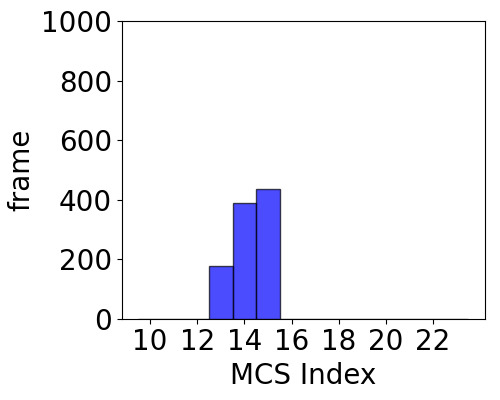}
        \caption{}
        \label{fig:rdm}
    \end{subfigure}
    \begin{subfigure}[b]{0.15\textwidth}
    \centering
        \includegraphics[width=\textwidth]{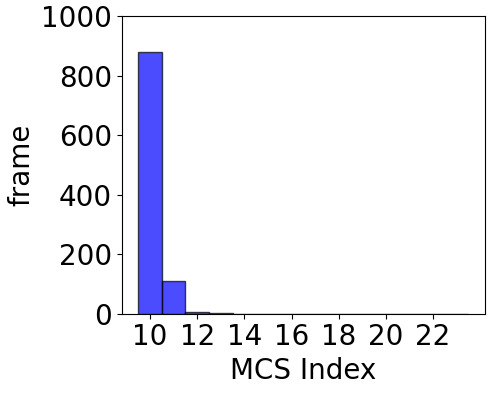}
        \caption{}
        \label{fig:rdm}
    \end{subfigure}
    \begin{subfigure}[b]{0.15\textwidth}
    \centering
        \includegraphics[width=\textwidth]{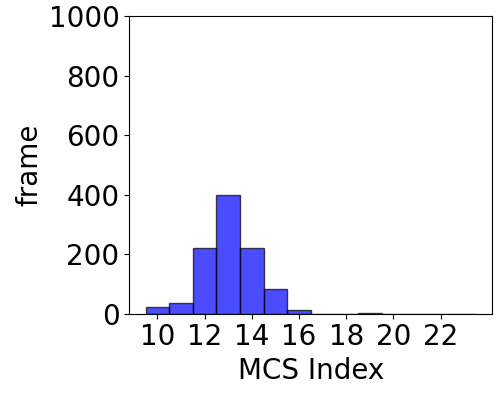}
        \caption{}
        \label{fig:rdm}
    \end{subfigure}
    \caption{Optimal MCS histogram in different channel scenarios and modes of UMi-LoS channel model: (a) static one user cluster (b) static two user clusters (c) static uncorrelated mode (d) static random placement (e) mobile one user cluster, and (f) mobile random placement.} 
    \label{fig:his}
\end{figure}

To generate training data, we configure QuaDRiGa software to simulate a 32 (base station antennas) $\times$ 4 (single-antenna UE) MIMO channel. In the model, the base station is positioned at the center of a circular area with a radius of 100 meters. For dataset completeness, we generate channels under 3D Urban Micro (UMi) Line-of-Sight (LoS), UMi Non-Line-of-Sight (NLoS), Urban Macro (UMa) Los, and UMa NLoS models. For each, we consider two channel scenarios: static and mobile. For static scenarios, we conduct four different modes: 1) uncorrelated mode, where users are positioned at considerable distances from each other, 2) one user cluster where all users share the same scatterers and hence have highly correlated channels, 3) two user cluster with two users in each cluster, and 4) random user placement. In the mobile scenario, users within this circle move in various directions at different speeds, with an average speed of 2.8 m/s. We also consider two distinct initial user placements in the mobile scenario: 1) co-located, and 2) randomly placed. In the co-located case, the users have high initial inter-user correlation and correlation will be decreased gradually as users move to different directions. In the random initial placement, more diverse channel conditions are covered. In total, we generated channel matrices for 13.5K transmission frames to be replayed into the emulated massive MIMO network environment by the Agora software. We utilized the MCS Table 2 of the 3rd Generation Partnership Project (3GPP) Technical Specification 38.214~\cite{3gpp.38.214} with modulation schemes, including 4-QAM, 16-QAM, 64-QAM, and 256-QAM, as well as LDPC code rates ranging from 0.11 to 0.92. In our study, we only considered MCS indices ranging from 10 to 24 in the table. This restriction arises from the impracticality of adopting MCS values that are either too high or too low, as they lead to intolerable bit error rates or inadequate spectral efficiency. Additionally, we adopt the Bit Error Rate (BER) as our performance metric. We collected datasets for per-frame BER for each channel model. Specifically, we conduct multiple rounds of evaluation per channel, exploring all available MCS indices. The optimal MCS for a given channel scenario is then defined as the MCS index that achieves an acceptable BER (i.e., lower than $10^{-3}$~\cite{ber}) in the largest number of frames. In Fig.\ref{fig:his}, we present the histogram depicting the distribution of optimal MCS across various channel scenarios and modes in UMi-LoS channel model, encompassing 1000 frames. Notably, the frequency distribution of optimal MCS aligns closely with the results of inter-user correlation analysis. Specifically, in static scenarios with one user cluster mode, high inter-user correlation is encountered due to the proximity of users. Consequently, the lowest MCS (10) is optimal as it ensures an acceptable BER in most cases. In contrast, ``two user clusters" mode exhibits less inter-user interference than the one user cluster but more than the uncorrelated mode and random placement. When transitioning to a mobile scenario, the correlation changes more randomly due to user mobility, resulting in a more dispersed distribution of optimal MCS over the 1000 test frames.

\begin{figure}[t]
    \centering
    \begin{subfigure}[b]{0.24\textwidth}
    \centering
      \includegraphics[width=1.72in]{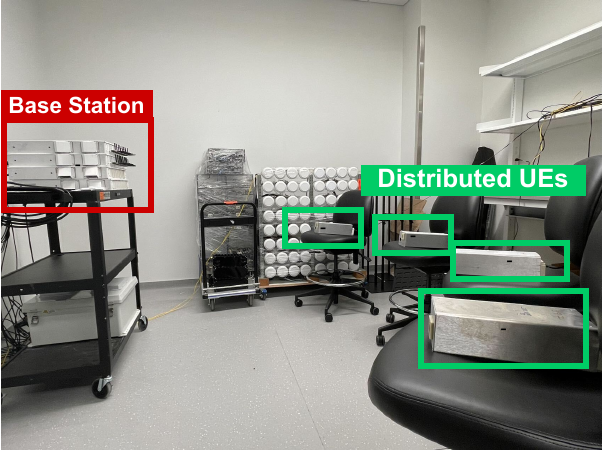}
      \caption{}
      \label{fig:exp1}
    \end{subfigure}
    \hfill
    \begin{subfigure}[b]{0.24\textwidth}
    \centering
      \includegraphics[width=1.72in]{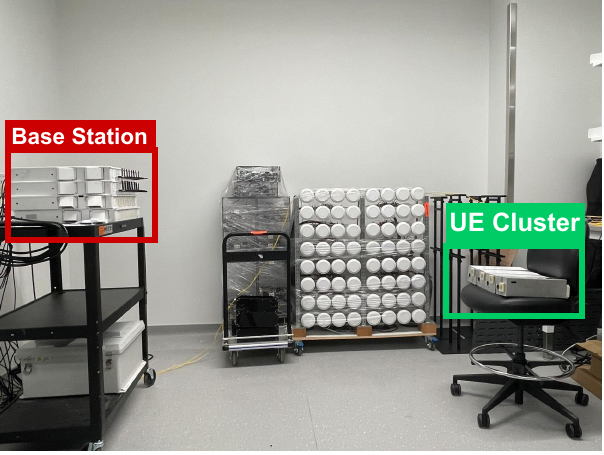}
      \caption{}
      \label{fig:exp2}
    \end{subfigure}
        \begin{subfigure}[b]{0.24\textwidth}
    \centering
        \includegraphics[width=1.72in]{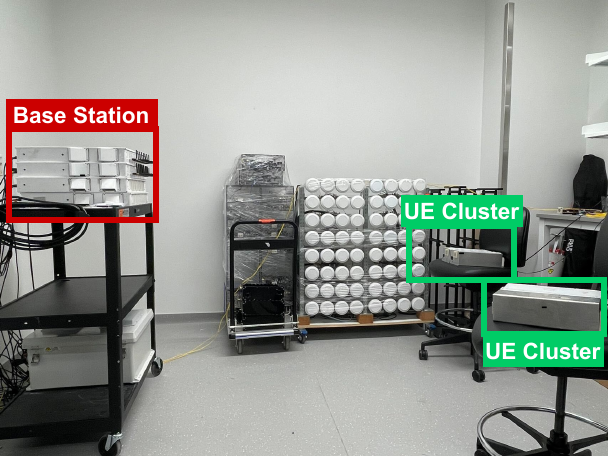}
        \caption{}
        \label{fig:exp3}
    \end{subfigure}
    \hfill
    \begin{subfigure}[b]{0.24\textwidth}
    \centering
        \includegraphics[width=1.72in]{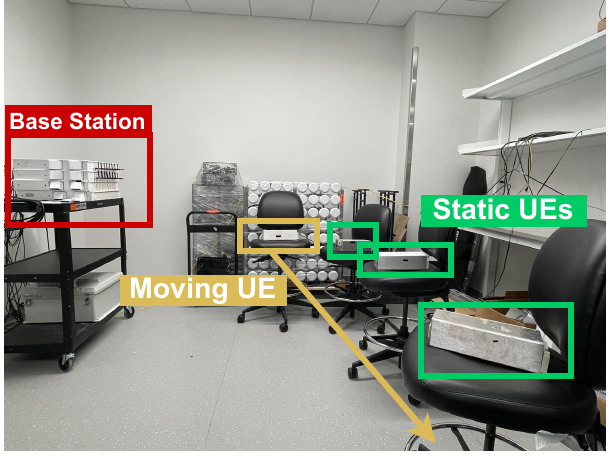}
        \caption{}
        \label{fig:exp4}
    \end{subfigure}
    \caption{OTA experiment deployment in static scenario: (a) uncorrelated mode (b) one user cluster and (c) two user clusters, and (d) user mobility scenario.} 
    \label{fig:OTA}
\end{figure}

We also conduct OTA experiments on the RENEW Massive MIMO platform~\cite{renew} to collect real-world datasets using Agora. Fig.~\ref{fig:OTA} illustrates our OTA experiment setup. We consider two channel scenarios (i.e. static and mobile). For the static scenario, uncorrelated, one-user cluster and two-user clusters are emulated. In the mobile scenario, we maintained three fixed users while moving the fourth user towards the static ones in a sequential manner. Throughout this process, the inter-user correlations between the mobile user and the static users varied with the changing distance between each static user and the mobile user. 


We use both simulated and real channel datasets to train our ML model presented in~\S\ref{sec:alg}. We run our model training on an NVIDIA DGX A100 server~\cite{nvidiadgx}. For DenseNet, we use three interleaved dense blocks and pooling layers. Each dense block comprises four 2-D convolution layers, two concatenation layers, and a batch-normalization (BN) layer. A flattening layer is utilized to process the CNN output before passing it to the LSTM. The LSTM module features three memory blocks, each with four interacting layers. We used PyTorch~\cite{pytorch} and the SGD optimizer~\cite{sgd} to train the model. Relevant simulation parameters are summarized in Table I.

\begin{table}[]
\centering
\caption{Experiment Setting and Training Parameters}

\begin{tabular}{cllcl}

\hline
\multicolumn{3}{c}{\textbf{Parameter}}       & \multicolumn{2}{c}{\textbf{Value}}     \\ \hline
\multicolumn{3}{c}{System Bandwidth}         & \multicolumn{2}{c}{20 MHz}             \\
\multicolumn{3}{c}{System Carrier Frequency} & \multicolumn{2}{c}{3.6 GHz}            \\
\multicolumn{3}{c}{Frame Duration}             & \multicolumn{2}{c}{1 ms}             \\
\multicolumn{3}{c}{MCS Table}             & \multicolumn{2}{c}{MCS index table 2 for PDSCH\cite{3gpp16p2}}                \\
\multicolumn{3}{c}{Cell Radius}              & \multicolumn{2}{c}{100 m}              \\
\multicolumn{3}{c}{UE Speed}                 & \multicolumn{2}{c}{0 \& 2.8 m/s}            \\
\multicolumn{3}{c}{Number of BS Antennas}    & \multicolumn{2}{c}{32}             \\
\multicolumn{3}{c}{Number of UEs}            & \multicolumn{2}{c}{4}             \\
\multicolumn{3}{c}{Batch Size}               & \multicolumn{2}{c}{64}                \\
\multicolumn{3}{c}{Learning Rate}      & \multicolumn{2}{c}{1e-3}               \\
\multicolumn{3}{c}{Optimizer}                 & \multicolumn{2}{c}{SGD}               \\
\multicolumn{3}{c}{Episodes}                 & \multicolumn{2}{c}{300}                \\
\hline
\end{tabular}
\label{tb1}
\end{table} 

\subsection{Benchmarks}

To comprehensively compare AMC techniques, we employ state-of-the-art DRL and Look-Up-Table (LUT) methods as well as a CNN-only method. 

\textbf{DRL-based AMC:} Deep reinforcement learning is favored in AMC for its ability to learn autonomously from environmental interactions, obviating the need for pre-collected datasets. Accordingly, we employ a Deep Q-Network (DQN) approach, as demonstrated in prior works~\cite{10024770, 9906093}. To maintain the historical MCS selection context, we expand the state representation by including not only the current frame's channel matrix but also those of the last two frames. The DQN model generates actions representing the chosen MCS index. The reward function assigns a value of 100 if the action matches the pre-collected optimal MCS index, and 0 otherwise. For our neural network architecture, we employ a five-layer structure, each with 64 neurons, and fine-tune hyperparameters via grid search to optimize performance.

\textbf{LUT AMC:} The LUT method is a traditional technique employed in the 5G standard. Nevertheless, its performance is significantly impacted by factors such as 5G numerology, the number of user spatial streams, propagation conditions, and traffic characteristics. In~\cite{map}, a novel Signal-to-Noise Ratio (SNR)-CQI-MCS mapping table is introduced and assessed within the context of 5G multi-user scenarios encompassing audio, video, and gaming traffic patterns, achieving a performance enhancement of approximately $35\%$ compared to state-of-the-art mapping tables. Consequently, we implement this innovative LUT table and conduct a comparative analysis with our proposed approach.

\textbf{CNN-based AMC:} The last benchmark we conducted is for an ablation study where we disabled the LSTM component in our proposed model, relying solely on CNN for prediction. As depicted in Sec.~\ref{sec:alg}, sequential dependencies are pivotal in the context of AMC. Previous frame channel conditions offer valuable insights for selecting the MCS in the current frame. Disabling the LSTM responsible for extracting inter-frame sequential channel information would compromise prediction accuracy.

\subsection{Experiment Results}
\subsubsection{Training and Testing Results}
We gathered a comprehensive dataset consisting of 16k simulation and OTA samples, encompassing diverse channel scenarios and modes. This dataset was partitioned into 13.5k samples for training and 2.5k for testing purposes. Our model underwent training for 300 epochs, resulting in progressively improving training and testing accuracy, as depicted in Fig.~\ref{fig:acc}. Ultimately, we achieved a training accuracy of $97.6\%$ and a testing accuracy of $92.5\%$.


\subsubsection{Performance Results}

\begin{figure}[t]
    \centering
    \begin{subfigure}[b]{0.24\textwidth}
    \centering
      \includegraphics[width=\textwidth]{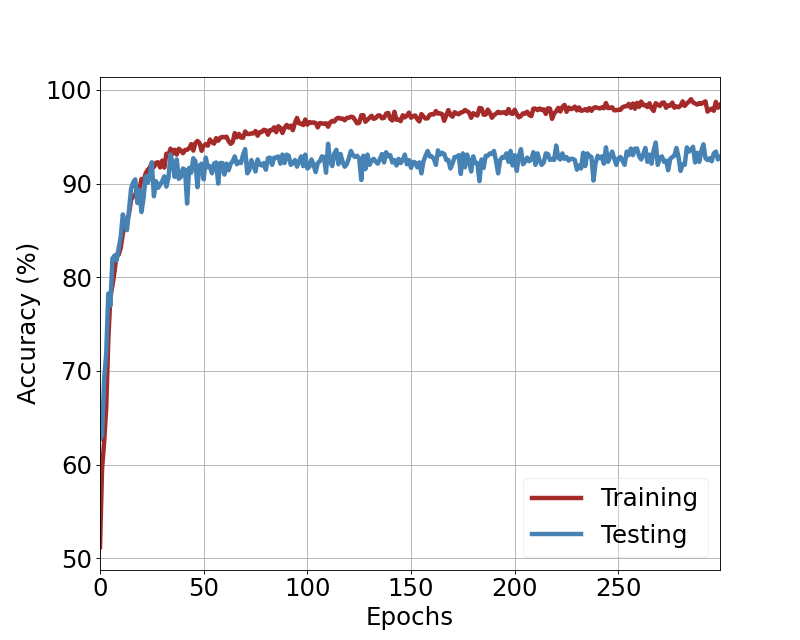}
      \caption{}
      \label{fig:acc}
    \end{subfigure}
    \hfill
    \begin{subfigure}[b]{0.24\textwidth}
    \centering
    \includegraphics[width=\textwidth]{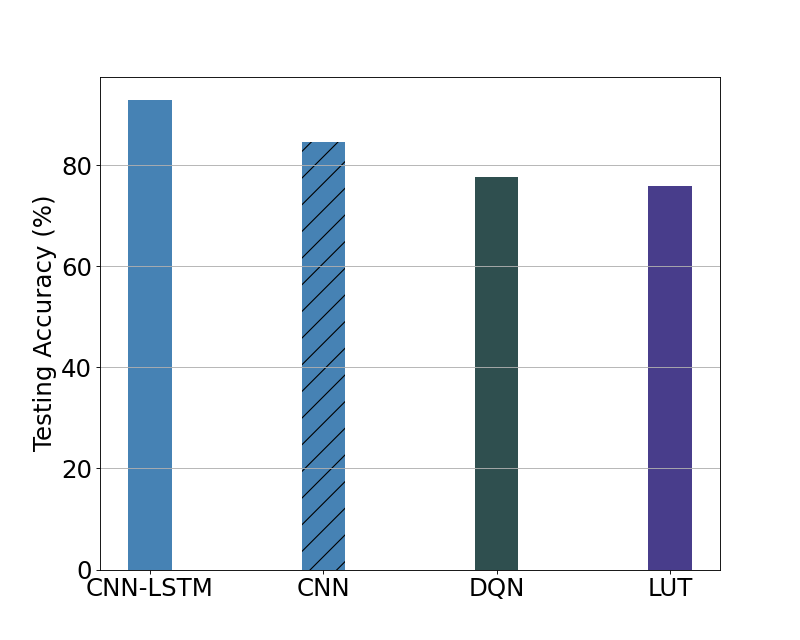}
      \caption{}
      \label{fig:comp}
    \end{subfigure}
    \caption{(a) Training and testing accuracy during training process and (b) Testing accuracy comparison} 
    \label{fig:perf}
\end{figure}

We evaluate the testing accuracy of our CNN-LSTM model in comparison to other benchmark models. For CNN and DQN-based approaches, we employ the testing dataset for inference and compare their predictions to pre-collected optimal MCS indices. In the case of the Look-Up Table (LUT), we map recorded SNR to CQI and then to MCS indices for comparison with the optimal MCS indices, assessing testing accuracy. Our comparison results are presented in Fig.~\ref{fig:comp}. The CNN-LSTM model outperforms other benchmarks due to its ability to effectively extract sequential and spatial channel information. Additionally, DQN performs less effectively in terms of testing accuracy, primarily because it struggles with the high-dimensional raw channel matrix from a 32-antenna base station and 4 UEs. Besides, the LUT exhibits the poorest performance among the benchmark models due to its limited adaptability to diverse channel conditions. Notably, the incorporation of LSTM into our model yields an impressive $8\%$ accuracy improvement compared to the CNN-only model. 


\section{Conclusion}
\label{sec:con}
In this paper, we presented a CNN-LSTM-based adaptive modulation and coding (AMC) technique for massive MIMO networks. Our method relies only on the channel state information of the users to predict the modulation and coding index (MCS) in MU-MIMO transmissions. To train and evaluate our model, we acquire both simulated and real-world data from the Agora PHY/MAC software and the RENEW massive MIMO platform under a variety of channel conditions including user mobility. Our findings reveal superior performance over existing state-of-the-art massive MIMO AMC methods across diverse channel scenarios and modes. 

\bibliographystyle{unsrt}
\bibliography{ref}

\end{document}